\documentclass[twocolumn,trackchanges]{aastex631}
\usepackage{natbib}
\usepackage{epsfig}
\usepackage{graphicx}
\usepackage{subfigure}
\usepackage{float}
\usepackage{amsmath}
\usepackage{color}
\usepackage{amssymb}
\usepackage{amsfonts}
\usepackage{apjfonts}

\begin{document}

\title{Galaxy Ages with Redshift $z=2-4$: Stellar Population Synthesis for Candidates in FourStar Galaxy
Evolution Survey}

\correspondingauthor{$\text{
Mart\'{i}n L\'{o}pez-Corredoira}$}
\email{martin@lopez-corredoira.com}

\correspondingauthor{Jun-Jie Wei}
\email{jjwei@pmo.ac.cn}

\author[0000-0003-0593-936X]{Chong-Yu Gao}
\affiliation{Purple Mountain Observatory, Chinese Academy of Sciences, Nanjing 210023, China}
\affiliation{School of Astronomy and Space Sciences, University of Science and Technology of China, Hefei 230026, China}

\author[0000-0001-6128-6274]{$\text{
Mart\'{i}n L\'{o}pez-Corredoira}$}
\affiliation{Instituto de Astrof\'{i}sica de Canarias, E-38205 La Laguna, Tenerife, Spain}
\affiliation{PIFI-Visiting Scientist 2023 of China Academy of Sciences at Purple Mountain Observatory, Nanjing 210023 and National Astronomical Observatories, Beijing 100101, China}
\affiliation{Departamento de Astrof\'{i}sica, Universidad de La Laguna, E-38206 La Laguna, Tenerife, Spain}

\author[0000-0003-0162-2488]{Jun-Jie Wei}
\affiliation{Purple Mountain Observatory, Chinese Academy of Sciences, Nanjing 210023, China}
\affiliation{School of Astronomy and Space Sciences, University of Science and Technology of China, Hefei 230026, China}



\begin{abstract}
Observations of large amounts of massive galaxies with relatively old populations found at high redshifts
are challenging galaxy formation scenarios within standard
cosmology. Precise determinations of the average age of these galaxies would be useful for the discussion of this problem.
Here we carry out a better constraint of
the age of 200 V-shaped spectral energy distribution (SED) nonactive galactic galactic nucleus galaxies at redshifts $2<z<4$ of the catalog of the FourStar Galaxy
Evolution Survey, identified by the V shape in their SED
with a Lyman and a Balmer break. The SED fitting
includes a main stellar population in addition to a residual younger population and extinction.
The galaxies are younger at a higher redshift on average. However,
for the galaxies with $z>2.5$, we do not see a significant evolution of their average age, with all average ages of the
galaxies mostly remaining between 1 and 2 Gyr. Our research finds that most massive galaxies ($\sim 10^{10} M_\odot$ ) are older (typically $>\sim 1$ Gyr old) and formed earlier than less massive galaxies in our sample.
\end{abstract}



\section{Introduction} \label{sec:intro}

Massive bright galaxies with relatively old populations have been found at high redshifts \citep[e.g.,][]{van_2001,van_2004,McCarthy_2004,
cimatti2004old,saracco2005density,Rodighiero2007,Cas12,Rie13,m2017,labbe2023}.
The formation of a galaxy requires a considerable amount of time. This fact and other observations, such as the existence of 
distant quasi-stellar objects (QSOs) at $z=7.6$ with black hole mass larger than $\mathrm{10^9}$ ${\rm M}_{\odot}$ \citep[e.g.,][]{Wang2021}, or the presence of high metallicity at high redshifts \citep[e.g.,][]{Becker_2001,Fin13}, or large amounts of molecular gas \citep[e.g.,]{Wal03,Nee20}, challenge the model of galaxy formation within the standard cosmological model.

The determination of the age of the galaxies is very important to evaluate how challenging are these high-redshift ($z\gtrsim 2$) massive galaxies are, either
with analysis of spectra or the spectral energy distribution (SED) from the photometry of galaxies in several filters.
Spectral analyses may give more accurate measurements but require very large exposure times, so they are available only for few galaxies,
and they are importantly affected by age-metallicity degeneracy in low signal/noise cases \citep{Car03}.
Nonetheless, there are useful studies in this direction.
For instance, \cite{Spi97} used the breaks of 2640 \AA \ and 2900 \AA \ and the
fit of the whole spectrum to the models to determine
the age of a galaxy at $z=1.55$ to be larger than 3.5 Gyr.
\citet{Sch06} used the strength of H$_\delta $ for $z\sim 0.9$ galaxies and the
fit of the whole spectrum to the models. Their method seems also
appropriate for our galaxies with redshift up to 1.2.
A more accurate method of age determination, such as the
use of H$_\gamma $ \citep{Vaz99,Yam06} would need a very high signal/noise in the spectra
which is not reachable for high-redshift galaxies.
However, \citet{Gla17} and \citet{Sch18} were able to analyze the spectra of quiescent galaxies from the FourStar Galaxy Evolution Survey (ZFOURGE)
using combined H$_\beta$+H$_\gamma$+H$_\delta $
absorption lines at $3<z<4$, getting stellar ages $\lesssim 1$ Gyr.

The stellar population synthesis (SPS) method has been used in exploring the age and other properties from the SED of galaxies. There
are several examples in the literature with remarkably high ages in comparison with the age of the Universe at the corresponding redshift.
\citet{Labbe2005} found three old galaxies with luminosity-weighted ages of 2.6 Gyr at $z=2.7$, 3.5 Gyr at $z=2.3$, and another 3.5 Gyr old galaxy at $z=2.3$;
\cite{Toft2005} used the SPS method to obtain the red compact galaxies with ages of 5.5, 3.5, and 1.7 Gyr for redshifts 1.21, 1.9, and 3.4, respectively.
At much higher redshift, $z\approx 6$, in 13 spectroscopically confirmed star-forming galaxies, the median
best-fitting age of the SED was 200--300 Myr \citep{Cur13}; a galaxy with photometric redshift 6.6--6.8
had an age $\gtrsim 50$ Myr and most likely a value of a few hundred Myr \citep{Ega05};
or a galaxy at $z=8.6$ had an age 450-500 Myr and a galaxy at $z=11.1$ with $\sim 160$ Myr \citep{Roc18}.
\citet{Vag22} provided a compilation of ages of 61 galaxies with redshifts up to $z\lesssim 4$, using Cosmic Assembly Near-infrared Deep Extragalactic Legacy Survey
(CANDELS) photometry for the most distant sources.

These SED fits are mostly carried out assuming one single stellar population (SSP) in the SPS, that is, assuming a passive evolving population;
and constraining the age to be lower than the age of the Universe at a given redshift.
\citet{m2017} claimed that by assuming only one SSP the age of galaxies would be underestimated.
This is a problem both for SED fitting or by deriving ages from spectral features
such as a breaks or some lines.
\citet{m2017} proposed two-SSP-fitting method (containing two stellar population components) to fit the SED of galaxies at $z>2.5$.
The young component of this double SSP, less than 5\% in mass, is consistent with a residual star formation that can also be observed at lower-redshift galaxies.
Their results indeed show that the two-SSP-fitting method can describe the SED of their candidates much better than one SSP,
but the uncertainties of the ages of older stellar components are high, so many galaxies should be analyzed to get a
good constraint in the average age for the observed redshift.

In this paper, we want to contribute to a better constraint of the age of red galaxies at redshifts $2<z<4$ by analyzing more than 200 galaxies
of ZFOURGE with SED fitting using the two-SSP method to derive the ages of young stellar components, ages of old stellar components, metallicities, redshift,
the ratio of old/young population, and extinction parameters. In our fit, we do not force the fit to be younger than the age of the Universe at their corresponding redshift. The objective of this analysis is
characterizing galaxies with V-shaped photometric SED at intermediate redshift, which will be useful in the comparison with other analyses of other higher redshift galaxies
selected with the same criteria that are being observed with the James Webb Space Telescope (JWST) \citep{Lop24}.
Once we have the fit for each galaxy,  we can calculate the average age of the galaxies
in the same bin, with a much lower error bar.

This article is organized as follows: In section \ref{sec:data} we describe the data provided by the ZFOURGE catalog and the selection criteria we used. In section \ref{sec:method}, the fitting model and the method to calculate the average age for each redshift bin are introduced. Our results are shown in section \ref{sec:result}. Finally, a brief discussion and conclusion are made in section \ref{sec:dis}.

\section{Data} \label{sec:data}
\subsection{ZFOURGE} \label{subsec:dat}

ZFOURGE provided a photometric catalog containing more than 70,000 galaxies, which covers the cosmic fields CDFS, COSMOS, and UDS. The observation of the galaxies in the ZFOURGE catalog involves numerous bandpass filters, with 40 filters for the CDFS field, 37 for the COSMOS field, and 26 for the UDS field. The wavelength extends from $\sim0.4\;\mu $m to $\sim7.9\;\mu $m \citep{Straatman2016}. In this catalog,
the information of each source include the identification number, R.A., decl., and flux (in units of $\mu $Jy) for each filter, along with its uncertainty and weight, signal-to-noise ratio (SNR), photometric redshift, spectroscopic redshift, and active galactic nucleus (AGN) status. The photometric redshift is obtained by fitting the spectral energy densities with the EAZY package \citep{Brammer_2008}.
In this work, we mainly use the flux for each filte, SNR, and AGN status to select our candidates and perform the further analysis.

\subsection{Sample selection} \label{subsec:sampleselect}

\cite{labbe2023} applied the empirical selection criteria to select high-redshift galaxies based on JWST-NIRCam photometry. They selected
on a `double-break' SED: no detection in HST-ACS imaging (with $\mathrm{SNR}(B_{435},V_{606},I_{814})<2$), blue in the NIRCam short-wavelength filters, and red in the NIRCam long-wavelength filters, which is expected for those sources at $z\ge7$ with a Lyman break and with red UV-optical colors. The adopted color selection criteria were
\begin{equation}
\label{criteria:labbe}
\begin{array}{l}
\text{F150W}-\text{F277W}<0.7 \\
\text{F277W}-\text{F444W}>1.0 .
\end{array}
\end{equation}
To ensure a good SNR, they required SNR(F444W)$>8$ and limited their sample to F444W$<27$ AB magnitude and F150W$<29$ AB magnitude. However, they are designed for $7 \leq z \leq 9$, and we cannot apply 
them directly. 

With the photometric redshifts provided by the ZFOURGE catalog, we first calculate the rest-frame wavelength associated to each filter. 
After that, we apply the color selection criteria
\begin{equation}
    \label{eq:select}
    \begin{array}{l}
    \text{FLyman} - \text{Fmed}<0.7 \\
    \text{Fmed}-\text{FBalmer}>1.0 ,
    \end{array}
\end{equation}
where `FLyman', `Fmed', and `FBalmer' are the AB magnitudes of the rest-frame central wavelengths of the filters, $\lambda_{f_{c}}$, 
such that $|\lambda_{f_{c}}-0.1216\;\mu m|<0.05\;\mu$m, $|\lambda_{f_{c}}-0.26\;\mu$m$|<0.05\;\mu$m, and 
$|\lambda_{f_{c}}-0.4\;\mu$m$|<0.05\;\mu$m, respectively.
We set the average flux of filters with central wavelength $<0.5\;\mu$m to be lower than the flux of FLyman,
which is an approximation of the criterion $\mathrm{SNR}(B_{435},V_{606},I_{814})<2$ in \cite{labbe2023}.
In addition, we also require the candidates have $\mathrm{SNR}>8$ to ensure good quality.
Note that the AGN effects can be excluded by using the information from the catalog.

\begin{figure*}
\begin{center}
\vskip-0.1in
\includegraphics[width=0.45\textwidth]{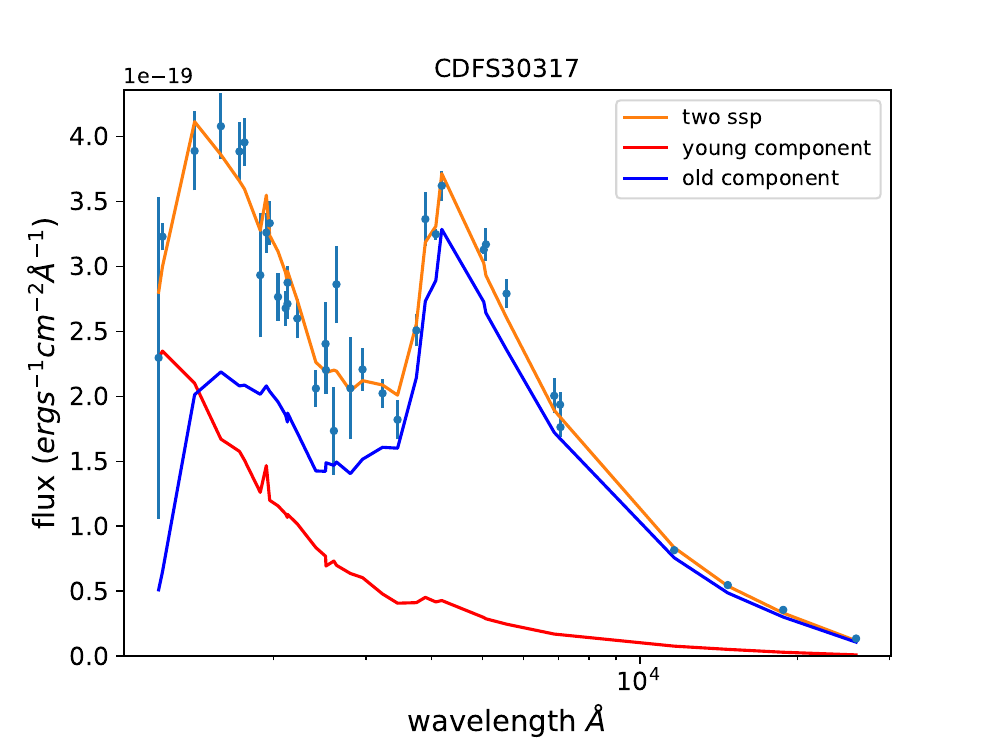}
\includegraphics[width=0.45\textwidth]{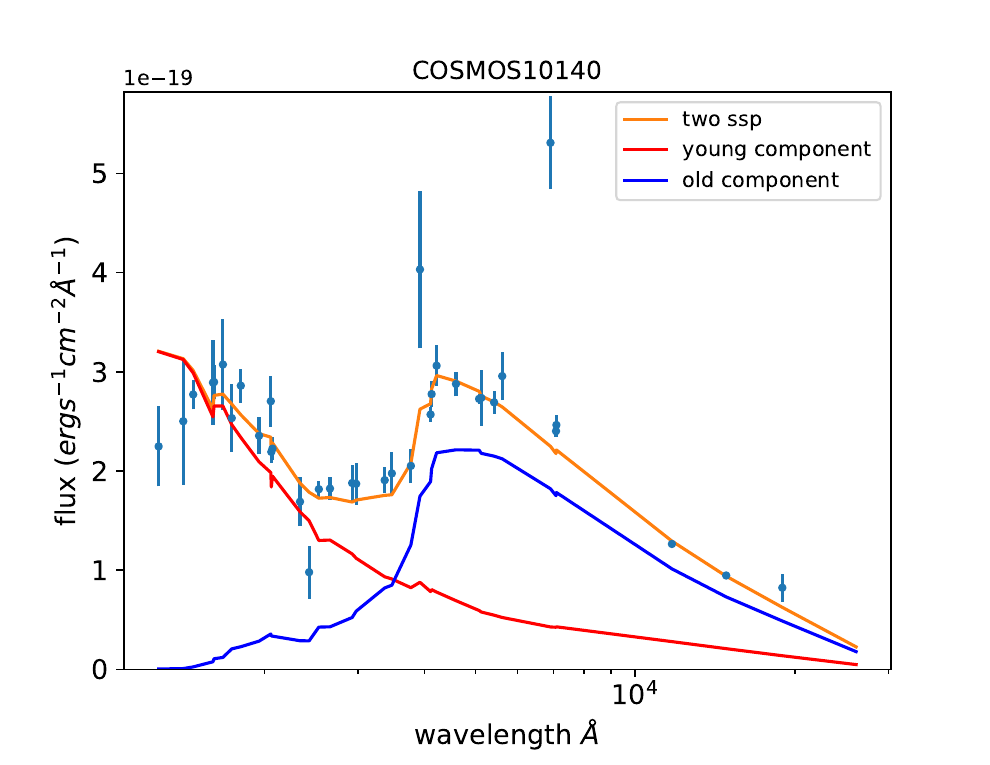}
\includegraphics[width=0.45\textwidth]{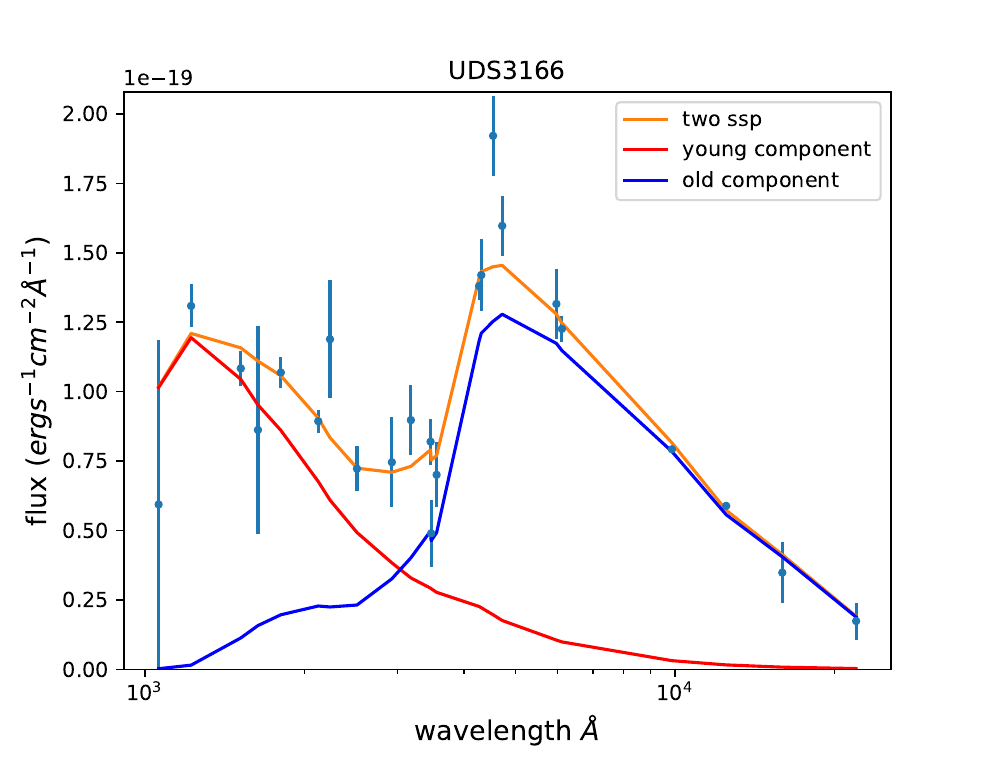}
\caption{Examples of the SED fitting by the two SSP method, shown here at rest assuming
the redshift inferred from the fitting. The best-fit ages of old stellar population are $0.29^{+0.35}_{-0.19}$, $0.64_{-0.35}^{+0.75}$, 
and $0.64^{+0.26}_{-0.35}$ Gyr, respectively. The best-fit ages of young stellar population are $0.025^{+0.075}_{-0.020}$, $0.005^{+0.020}_{-0.000}$, and $0.005^{+0.02}_{-0.00}$ Gyr, respectively. The inferred redshifts for these three examples are $2.05_{-0.01}^{+0.01}$, $2.36_{-0.09}^{+0.36}$, and $2.60_{-0.10}^{+0.10}$, respectively. And the stellar masses are $10^{11} M_{\odot}$, $10^9 M_{\odot}$, and $10^9 M_{\odot}$}
\label{fig:sedexample}
\vskip-0.2in
\end{center}
\end{figure*}

In Figure~\ref{fig:sedexample}, we show three representative examples of SEDs, one each for the CDFS, COSMOS, and UDS fields. 
In this plot, the rest-frame wavelength ($\lambda_{\rm rest}$) is calculated by $\lambda_{\rm rest}=\lambda_{\rm obs}/(1+z_{\rm fitted})$,
where $z_{\rm fitted}$ is the redshift inferred from the fitting (see below) and $\lambda_{\rm obs}$ is the observer-frame wavelength. 
We have 15 candidates for the CDFS field, 84 candidates for the COSMOS field, and 101 candidates for the UDS field. The total number 
is 200 candidates. All of them present a `V shape' in $F_{\lambda}$ versus $\lambda_{\rm obs}$ as they have `double-break' 
features (i.e., Lyman and Balmer breaks).

\begin{figure}
    \centering
    \includegraphics[width=0.45\textwidth]{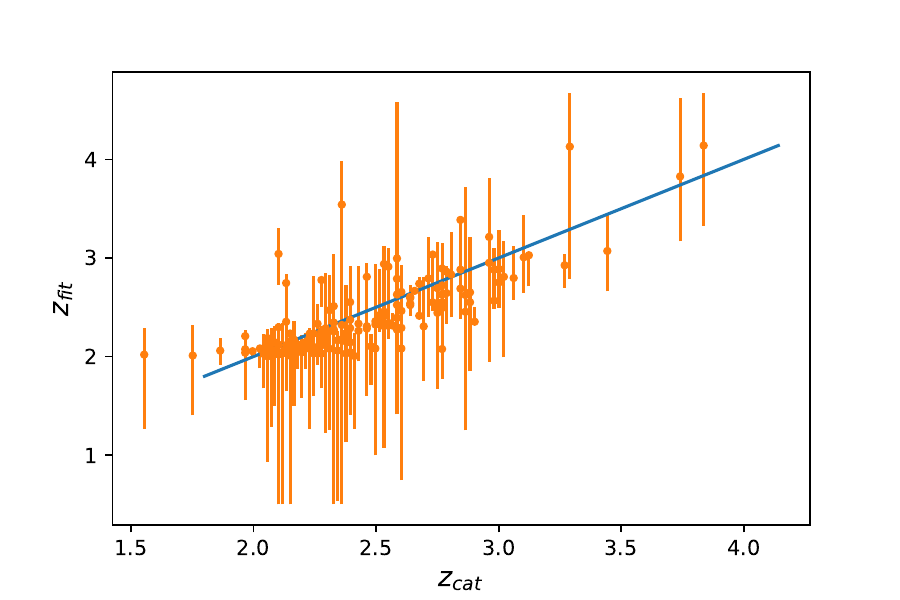}
    \caption{The comparison between the redshifts obtained by our method and those provided by the catalog.}
    \label{fig:zz}
\end{figure}

\section{Methods} \label{sec:method}

\subsection{Model fitting}
The physical properties of a galaxy can be obtained by fitting its SED with SPS models.
Most of the studies applied only one SSP, with one single average age, which is lower than the age of the oldest stellar population according to \cite{m2017}. It has been proved that the two-SSP method (containing a young and a relative old stellar population components) gives better analysis for the V-shaped candidates \citep{m2017}. Thus, in our analysis we use the two-SSP method to fit the SEDs of the candidates.

The SPS model we used is GALAXEV \citep{Bruzual2003}, which computes the spectral evolution of stellar populations 
from 91 \AA\ to 160 $\mu m$ at rest. GALAXEV contains ten ages (0.005, 0.025, 0.10, 0.29, 0.64, 0.90, 1.4, 2.5, 5.0, and 11 Gyr)
and three metallicities ([M/H]$=-0.4$, $0$, and $+0.4$), so there are 30 instantaneous-burst templates that can be used to fit the galaxy
spectra \citep{2003PhDT.........2T}. In this work, we use Calzetti's law to consider the extinction effects. Calzetti's law was empirically 
derived from a nearby starburst galaxies containing small Mgellanic cloud-like grains \citep{2000ApJ...533..682C}. This law is suitable for
galaxies at high redshifts, in which stars are formed within the central regions. Here the adopted extinction parameter is $R_V=4.05$.

Our fitting model can be expressed by the following equation:
\begin{equation}\label{eq:mod}
\begin{split}
F_{\rm theo}(\lambda)=&\frac{L_0}{4\pi d_L(z)^2(1+z)}\times \\
	&[\langle L_{\rm SSP}({\rm age _{\rm old},[M/H]}, A_V;\lambda /(1+z))\rangle_T \\
        &+A_2\langle L_{\rm SSP}({\rm age _{\rm young},[M/H]}, A_V;\lambda /(1+z)) \rangle_T ]\;,
\end{split}
\end{equation}
where $d_{L}(z)$ is the luminosity distance, $\text{age}_{\text{old}}$, $\text{age}_{\text{young}}$, $[M/H]$, $A_v$, $A_2$, and $z$, 
respectively, represent the age of old stellar population, the age of young stellar population, metallicity, the extinction in V-filter, 
the ratio of young/old population, and redshift. The flux provided by the catalog is in units of $\mu Jy$, so we need to convert it into
$\mathrm{erg}$ $\mathrm{s^{-1}}$ $\mathrm{cm^{-2}}$ \AA$^{-1}$. The conversion is carried out by means of
$F_\lambda({\rm erg} \mathrm{~s}^{-1} \mathrm{~cm}^{-2}\;\text{\AA}^{-1})=2.998 \times 10^{-14} F_\nu(\mu Jy) \times \frac{\Delta \nu}{\Delta \lambda},$ where $\Delta \lambda=\frac{\left[\int d \lambda T(\lambda)\right]^2}{\int d \lambda T^2(\lambda)}$, $\Delta \nu=\frac{\left[\int d \nu T(\nu)\right]^2}{\int d \nu T^2(\nu)}$, and $T(\lambda)$ is the transmission curve for the corresponding filter.

The model fitting is carried out by minimizing the $\chi^2$,
\begin{equation}
    \label{eq:chi2}
    \chi_{\text {red }}^2=\frac{1}{N_{\text {dof }}} \sum_{i=1}^N \frac{\left[F_{\text {obs}}\left(\lambda_i\right)-F_{\text {theo }}\left(\lambda_i\right)\right]^2}{\sigma^2\left(\lambda_i\right)} \text {. }
\end{equation}
Here $N_{\rm dof}=N-f$, where $N$ is the number of data points and $f$ is the number of free parameters ($f=7$ in our case).
Also, $\sigma _i$ is the error of the observed flux $F_{\text {obs}}\left(\lambda_i\right)$.
Although the distance luminosity $d_L$ is estimated in flat $\mathrm{\Lambda CDM}$ model with cosmological parameters
$H_0=70$ $\mathrm{km\;Mpc^{-1}\;s^{-1}}$ and $\Omega _m=0.3$, the SED fitting is totally independent of the cosmological model, because
$d_L$ only affects the amplitude of the SED, not its shape.

After obtaining the best-fit parameters through the minimum $\chi^2$ statistic, the error bars of the parameters
can be derived through

\begin{equation}
    \label{eq:errorbars}
    \chi_{\text {red }}^2<\chi_{\text {red,min}}^2\left[1+\frac{f(N_{\text {dof }},CL)}{N_{\text {dof }}}\right]\;.
\end{equation}
where $f(N_{\text {dof }},CL)$ is a function of $N_{\rm dof}$ and the confidence level (CL) given in the corresponding
table of $\chi ^2$ statistics.
In cases with the minimum $\chi^2_{\rm red, min}\ne 1$, the errors would be underestimated or overestimated.
We thus add a factor $\chi^2_{\rm red, min}$ in the right-hand side of Equation~(\ref{eq:errorbars}).
This is equivalent to assume that our fit is a ``good fit'' (with $\chi^2_{\rm red, min}\approx 1$) and multiplying 
the error bars by some factor to get it.

Figure~\ref{fig:zz} displays the comparison between the redshifts inferred from our model fitting ($z_{\rm fit}$) 
and those provided by the catalog ($z_{\rm cat}$). One can see from this plot that they are in good agreement within the error bars.
\subsection{Stellar Mass Estimation}
\begin{figure}
\centering
\includegraphics[width=0.5\textwidth]{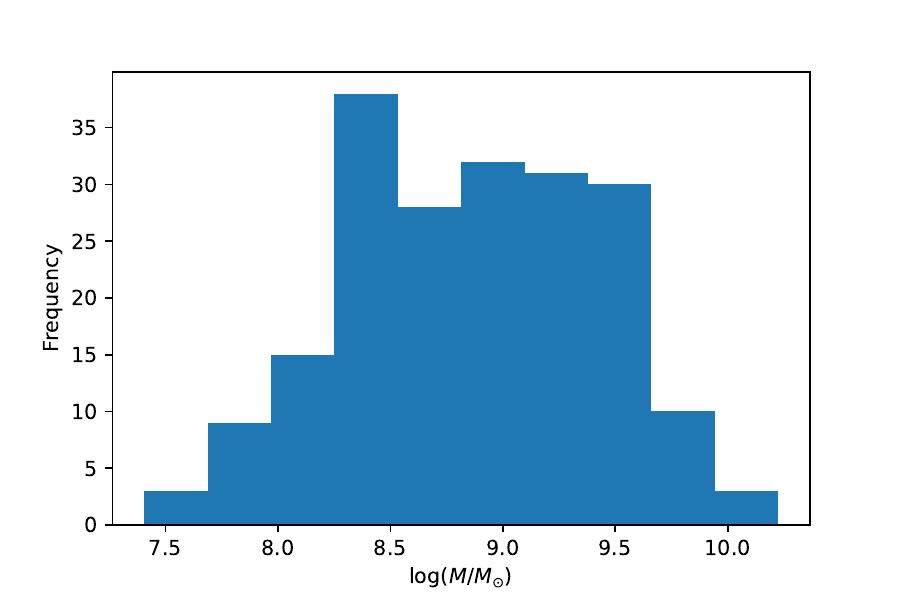}
\caption{ The stellar mass distribution of the 200 galaxies with redshift $z_{\rm fit}>2$.}
\label{fig:Mdistri}
\end{figure}

In our model, there are two stellar components in the galaxies. In order to estimate the total mass of a galaxy, each component should be 
considered, though we are dominated by the mass of the old population. 
We estimate the stellar mass of a galaxy by means of the following procedure: 
First, we calculate the rest-frame V-band luminosity $L_V$ by the rest-frame V-band flux (corrected of extinction). 
Then, according to equation \ref{eq:mod}, the young population contributes $L_{V,young} = L_V \times A_2/(1+A_2)$, and the old population contributes 
$L_{V,old} = L_V \times 1/(1+A_2)$. \cite{ml2005} calculated the stellar mass-to-light ratios with different ages by GALAXEV assuming the solar metallicities $\Upsilon _V(age)$,
which are used here to calculate the mass of the young and old population of a galaxy, neglecting the metallicity
dependence: $M=M_{young}+M_{old}$, where $M_{young}$ and $M_{old}$ represent the young and old components masses. That is, 
$M_{x}=L_{V,x} \times \Upsilon _V(age_{x})$, x could be old or young. The mass distribution of our datasets is shown in Fig. \ref{fig:Mdistri}. We note that the dispersion of masses shown in Fig. \ref{fig:Mdistri} is larger than real, because there is a larger error in the mass determination due to the errors of ages. The average mass of our sample of 200 galaxies is $1.5 \times 10 ^{9}$ $ M_{\odot}$, with range $10^{7}$ to $10^{11}$ $M_{\odot}$.

\subsection{Average and median ages of galaxies at different redshifts} \label{subsec:aver}
After obtaining the best-fit parameters $\mathrm{age_{old}}$ and their corresponding uncertainties for 727 galaxies,
We choose those galaxies with inferred redshift $z_{\rm fit}>2$, and end up with 200 galaxies.
Firstly, we divide our selected galaxies into 6 groups with redshifts from low to high, with each group containing the similar number of galaxies. 
The redshift bins are 2.00--2.05, 2.05--2.11, 2.11--2.23, 2.33--2.37, 2.37--2.66, and 2.66--4.14, respectively.

Assuming the positive and negative errors are Gaussian (an asymmetrical Gaussian when the positive and negative error are different),
the probability $P$ of each galaxy to have age $x$ is
\begin{equation}
\label{eq:assygauss}
\mathrm{P_i}(x)=\frac{\sqrt{2/\pi}}{\sigma_l+\sigma_r}\times \left\{\begin{array}{l}
\exp \left[-\frac{(x-\tau _i)^2}{2 \sigma_l^2}\right] \quad  {\rm if}\ x \leqslant \tau _i;\\
\exp \left[-\frac{(x-\tau _i)^2}{2 \sigma_r^2}\right] \quad  {\rm if}\ x>\tau _i.
\end{array}\right.
\end{equation}
where $\tau _i$ means the $i$th $\mathrm{age_{old}}$, $\sigma_l$ and $\sigma _r$, respectively, represents the negative
and positive $\mathrm{1 \sigma}$ errors.
By taking the cumulative product over all distributions, we can obtain the distribution of average age:
\begin{equation}
    \label{eq:prod}
    \mathrm{P_{aver}}(x)=K\times \prod _{i=1}^{n} \mathrm{P_{i}}(x) ,
\end{equation}
where $n$ is the number of galaxies in each redshift bin and $K$ is a normalization constant.
Afterward, by identifying the 0.16, 0.5, and 0.84 quantiles of the overall distribution,
we can obtain the lower, median, and upper limit of its $\mathrm{1 \sigma}$ confidence interval. The above calculation assumed that the galaxies in the same bins have similar ages, this underestimates the uncertainties of average ages. Considering the differences of the ages of galaxies in the same bin, the uncertainties should be multiplied a factor  $\sqrt{\frac{\chi^2_i}{n_i-1}}$, with $\chi ^{2}_{i} = \sum _{k=1}^{n_i} \frac{(log(age)_{k}-log(age)_{i,aver})^2}{\sigma _{log(age)_{k}}^2}$, where i stands for the ith bin, $n_i$ is the number of galaxies in the bin, $log(age)_{i,aver}$ is the average age of ith bin calculated above, and $\sigma _{log(age)_{k}}$ represents the log(age) uncertainties of kth galaxy in ith bin.

Median values are effected by the outliers more slightly, so we also calculate the median ages for each redshift bin and mass bin. The statistical error bars are calculated in 68\% C.L. ($\mathrm{1 \sigma}$ uncertainties): the upper or lower limits correspond to the positions $0.5N \pm 0.47\sqrt{N}$ of the ordered set of N data. Presenting results through average and median values could avoid anomalies caused by systematic errors in the fitting method.

\section{Results} \label{sec:result}

\begin{figure}
\centering
\includegraphics[width=0.5\textwidth]{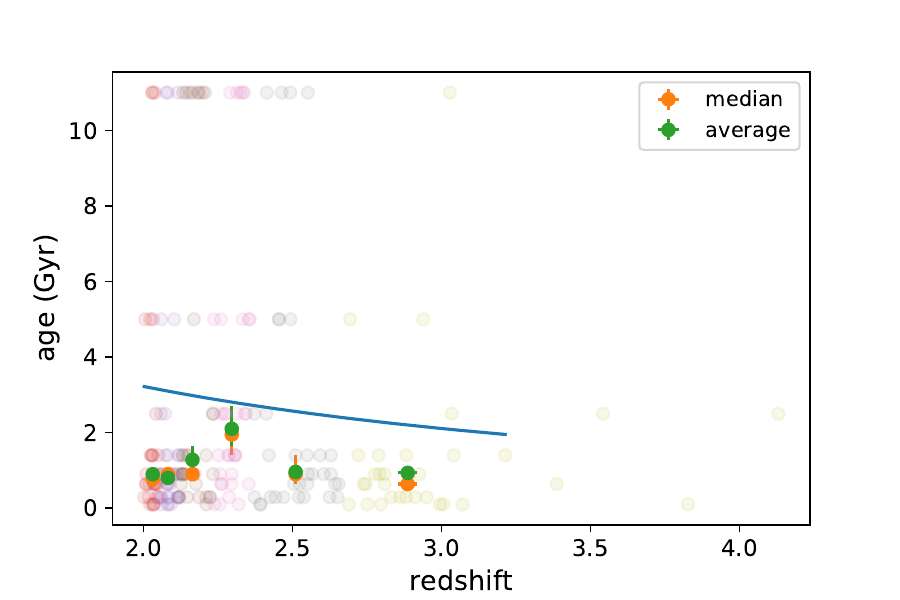}
\caption{The average and median ages of old stellar populations in 200 galaxies for 6 redshift bins. 
Vertical error bars correspond to the $1 \sigma$ confidence level of the average and median ages 
 and horizontal error bars indicating the standard deviation of the redshifts in such bin. The transparent dots represent the best-fitted ages of galaxies, and the colors of dots are used to distinguish between different bins. The main reason that some dots has ages higher than the universe age is that we did not plot the error bars of single galaxies ages in order to make the figure looks more concise, the error bars could be large.} The solid curve shows the theoretical age of the Universe as a function 
of redshift in ﬂat $\Lambda$CDM.
\label{fig:agez}
\end{figure}

\begin{figure}
\centering
\includegraphics[width=0.5\textwidth]{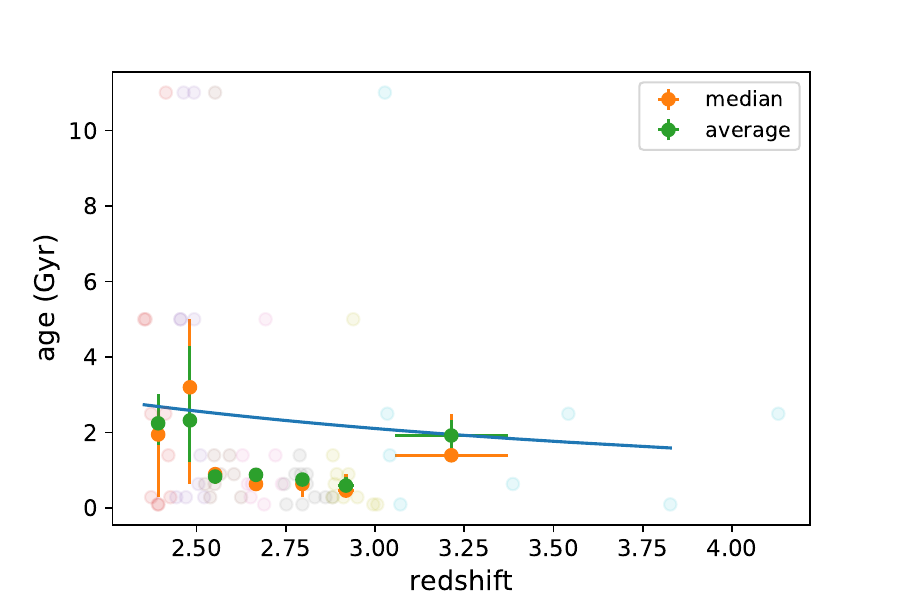}
\caption{Same as Figure~\ref{fig:agez}, but now for 70 highest-redshift galaxies.}
\label{fig:high}
\end{figure}

With the inferred ages of old stellar populations in 200 galaxies having $z_{\rm fit}>2$, we calculate the average age and median age of 
the galaxies for each redshift bin by using the method described in Section~\ref{subsec:aver}. Figure~\ref{fig:agez}
shows the age-redshift diagram for six redshift bins. In this plot, we also illustrate the age of the Universe as a function 
of redshift (estimated using flat $\Lambda$CDM; solid curve). It is obvious that the average and median galaxy ages are 
younger than the age of the Universe at the same redshift. One can find that the median and average galaxy ages are very similar in the figure \ref{fig:agez} to figure \ref{fig:high}. In order to study the correlation between the average age and $z_{\rm fit}$, 
we perform a linear fit, $t_{\rm gal}=A+B\times z_{\rm fit}$, to the data points. Here $t_{\rm gal}$ denotes the 
average age of the galaxies.The best-ﬁt coefficients are $A=0.62_{-0.42}^{+0.42}$ Gyr and $B=0.10_{-0.17}^{+0.18}$.
The slope $B$ is consistent with $0$ at the 1$\sigma$ conﬁdence level, implying that there is no significant correlation 
between $t_{\rm gal}$ and $z_{\rm fit}$.

\begin{figure}
\centering
\includegraphics[width=0.5\textwidth]{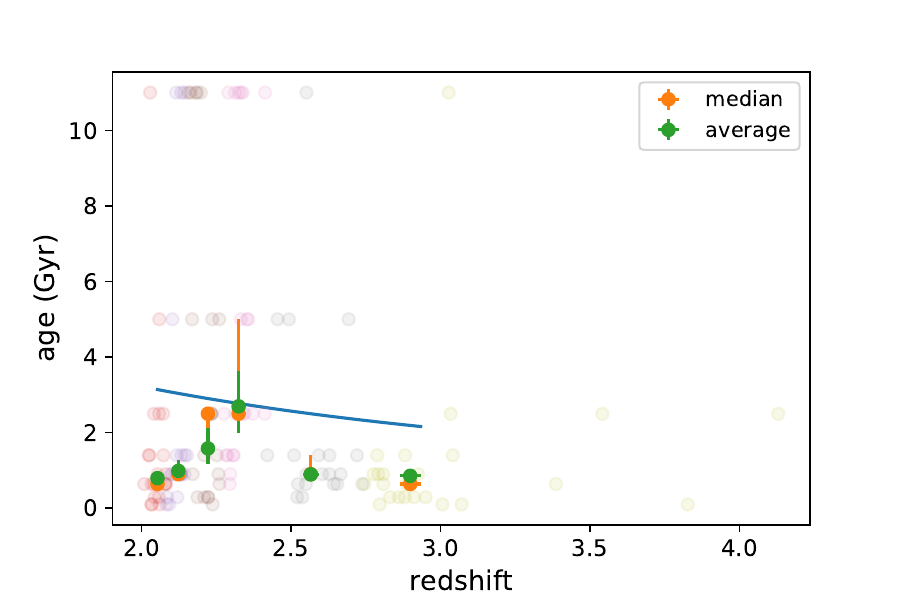}
\caption{Same as Figure~\ref{fig:agez}, but now for 130 galaxies with $L>5.75\times 10^{42}$ $\mathrm{erg\;s^{-1}}$.}
\label{fig:selection}
\end{figure}

There exist some selection effects such as the Malmquist bias, which leads to the preferential detection of galaxies
with higher luminosity. In order to check the impact of the Malmquist bias on our results, we will only consider 
those sufficiently bright galaxy sources with rest frame V-band luminosity $L>5.75\times 10^{42}$ $\mathrm{erg\;s^{-1}}$.  
Figure~\ref{fig:selection} shows the average ages of 130 galaxies with $L>5.75\times 10^{42}$ $\mathrm{erg\;s^{-1}}$
as a function of redshift. We can see that high-$L$ galaxies are observable even at high redshifts between 3 and 3.5.
Since there is no significant difference between Figures~\ref{fig:agez} and \ref{fig:selection}, we conclude that 
the selection effect is not important in this research. Similarly, a simple linear-fit analysis for the data points 
in Figure~\ref{fig:selection} gives $A=1.06_{-0.52}^{+0.52}$ Gyr and $B=-0.06_{-0.19}^{+0.19}$. The slope $B$ is also 
consistent with 0 at the 1$\sigma$ confidence level.

If we focus on 70 highest-redshift galaxies in our sample and divide them into 7 redshift bins, 
the range of $z_{\rm fit}$ is from 2.35 to 4.14. Their average galaxy ages are presented in Figure~\ref{fig:high}. 
We also perform a linear fit to the average age-redshift measurements, and obtain the results of 
$A=0.07^{+1.12}_{-1.12}$ Gyr and $B=0.28_{-0.42}^{+0.40}$. Again, the slope $B$ is in line with
0 at the 1$\sigma$ confidence level.

In Table \ref{tab:result}, we summarize the average galaxy ages in different redshift bins for the samples
of 200 galaxies, 130 galaixes with $L>5.75\times 10^{42}$ $\mathrm{erg\;s^{-1}}$, and 70 highest-redshift 
galaxies, respectively. In addition, the corresponding cosmic ages are also listed in Table \ref{tab:result}
for comparisons.

\begin{table*}
\centering
\caption{The Average and Median Galaxy Ages} and the Corresponding Cosmic ages in Different Redshift Bins for the Samples
of 200 Galaxies, 130 Galaixes with $L>5.75\times 10^{42}$ $\mathrm{erg\;s^{-1}}$, and 70 Highest-redshift 
Galaxies.
\label{tab:result}
\resizebox{\textwidth}{!}{
\begin{tabular}{cccccccccccc}
\hline
\multicolumn{4}{c}{200 Galaxies}             & \multicolumn{4}{c}{130 Galaxies with $L> 5.75\times 10^{42}$ $\mathrm{erg}$ $\mathrm{s^{-1}}$} & \multicolumn{4}{c}{70 Highest-redshift Galaxies}        \\ \hline
$z$               & Cosmic Age  & Average Galaxy Age & Median Galaxy Age  & $z$                           & Cosmic Age              & Average Galaxy Age   & Median Galaxy Age         & $z$               & Cosmic Age  & Average Galaxy Age  & Median Galaxy Age \\ 
               & (Gyr) & (Gyr) &  (Gyr)     &                      &  (Gyr)             & (Gyr)   &(Gyr)         &             & (Gyr) & (Gyr) & (Gyr)\\ 
\hline
$2.03 \pm 0.01$ & 3.17             & $0.90_{-0.15}^{+0.20}$ &$0.77_{-0.13}^{+0.13}$  & $2.04 \pm 0.02$             & 3.15                         & $0.79_{-0.14}^{+0.17}$   & $0.64_{-0.00}^{+0.26}$           & $2.39 \pm 0.02$ & 2.69            & $2.24_{-0.57}^{+0.77}$ & $1.95_{-1.66}^{+0.55}$  \\
$2.08 \pm 0.02$ & 3.08             & $0.80_{-0.06}^{+0.08}$  &$0.90_{-0.26}^{+0.00}$ & $2.11 \pm 0.02$             & 3.05                         & $0.98_{-0.22}^{+0.28}$        &$0.90_{0.00}^{+0.00}$      & $2.48 \pm 0.02$ & 2.59            & $2.32_{-1.09}^{+1.97}$ &$3.20_{-2.56}^{+1.80}$  \\
$2.16 \pm 0.04$ & 2.97             & $1.27_{-0.28}^{+0.37}$ &$0.90_{-0.00}^{+0.50}$  & $2.20 \pm 0.04$             & 2.92                         & $1.58_{-0.40}^{+0.54}$   &$2.50_{-1.10}^{+0.00}$           & $2.56 \pm 0.03$ & 2.50            & $0.83_{-0.09}^{+0.10}$ &$0.90_{-0.26}^{+0.00}$  \\
$2.29 \pm 0.03$ & 2.80             & $2.10_{-0.46}^{+0.60}$ &$1.95_{-0.55}^{+0.55}$  & $2.34 \pm 0.06$             & 2.75                         & $2.69_{-0.69}^{+0.94}$  &$2.50_{-0.00}^{+2.50}$            & $2.67 \pm 0.03$ & 2.38            & $0.88_{-0.12}^{+0.14}$  &$0.64_{-0.00}^{+0.26}$ \\
$2.51 \pm 0.08$ & 2.55             & $0.96_{-0.13}^{+0.17}$  &$0.90_{-0.26}^{+0.50}$ & $2.59 \pm 0.07$             & 2.46                         & $0.89_{-0.07}^{+0.08}$  &$0.90_{0.00}^{0.50}$            & $2.80 \pm 0.04$ & 2.27            & $0.75_{-0.08}^{+0.10}$  &$0.64_{-0.35}^{+0.26}$ \\
$2.98 \pm 0.31$ & 2.12             & $0.93_{-0.10}^{+0.11}$ &$0.64_{-0.00}^{0.26}$  & $3.00 \pm 0.30$             & 2.13                         & $0.85_{-0.08}^{+0.11}$         &$0.64_{-0.00}^{+0.26}$     & $2.92 \pm 0.04$ & 2.16            & $0.59_{-0.17}^{+0.23}$  &$0.46_{-0.17}^{+0.43}$ \\
         & &       &                  &                          &                             &                              &                                     & $3.36 \pm 0.37$ & 1.85            & $1.92_{-0.32}^{+0.41}$ &$1.40_{-0.00}^{1.10}$  \\ \hline
\end{tabular}}
\end{table*}

\begin{figure}
\centering
\includegraphics[width=0.5\textwidth]{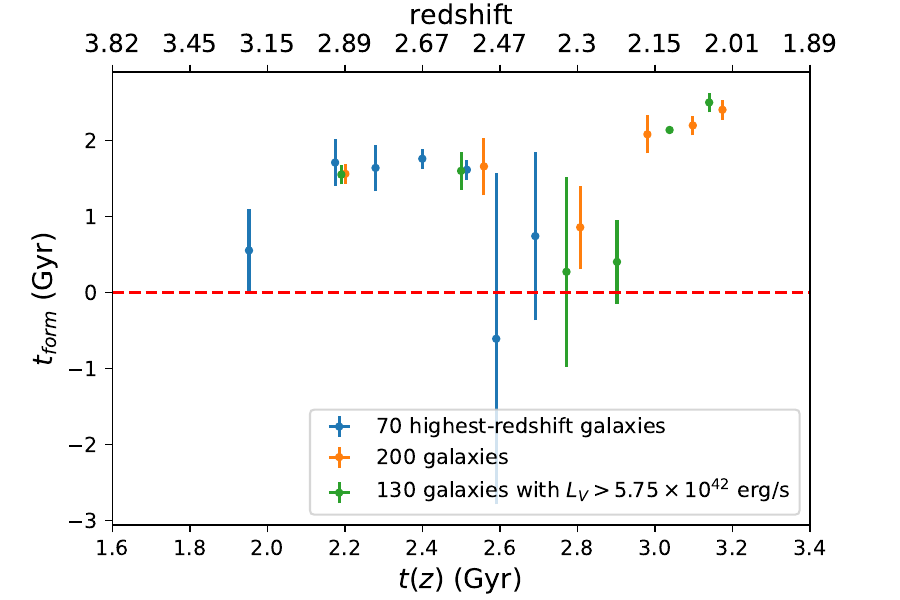}
\caption{The median galaxy formation times} as a function of cosmic time (or redshift) for the samples
of 200 galaxies (orange dots), 130 galaxies with $L>5.75\times 10^{42}$ $\mathrm{erg\;s^{-1}}$
(green dots), and 70 highest-redshift galaxies (blue dots), respectively. The red dashed line marks $t_{\rm form}=0$.
\label{fig:tform}
\end{figure}

We define the median formation time of the galaxy as $t_{\rm form}=t(z)-t_{\rm gal}$, 
where $t(z)$ is the age of the Universe at the corresponding redshift bin. Figure~\ref{fig:tform}
shows the median galaxy formation times as a function of cosmic time (or redshift) for the samples
of 200 galaxies (orange dots), 130 galaxies with $L>5.75\times 10^{42}$ $\mathrm{erg\;s^{-1}}$
(green dots), and 70 highest-redshift galaxies (blue dots), respectively. Except for three abnormal points,
the values of $t_{\rm form}$ range from $\sim 1$ Gyr to $\sim2.5$ Gyr, which are consistent with the results of \cite{Cas12}. 
Also, one can see from Figure~\ref{fig:tform} that for different galaxy samples, the average galaxy formation
times in similar redshift bins are similar, which means that the selection effects have minimal influences
on our results. Galaxy with stellar age of Y and stellar mass of X means the galaxy formed X mass in Y time. So the star formation rate of it is X/Y. For our result of figure \ref{fig:agez} to figure \ref{fig:high} we could estimate the average SFR for each bins by $\rm{SFR=\frac{<mass>}{<age>}}$, where $\rm{<mass>}$ and $\rm{<age>}$ are the average mass and average age of the galaxies in the correspond bin.The average SFR for each bin are calculated for Fig. \ref{fig:agez} to \ref{fig:high}, which fall in the range from $10^{-0.02}$ to $10^{0.7}$ $\rm{M_\odot / Yr}$, consist with the result from \cite{kouroumpatzakis2023star}.

\begin{figure}
\centering
\includegraphics[width=0.5\textwidth]{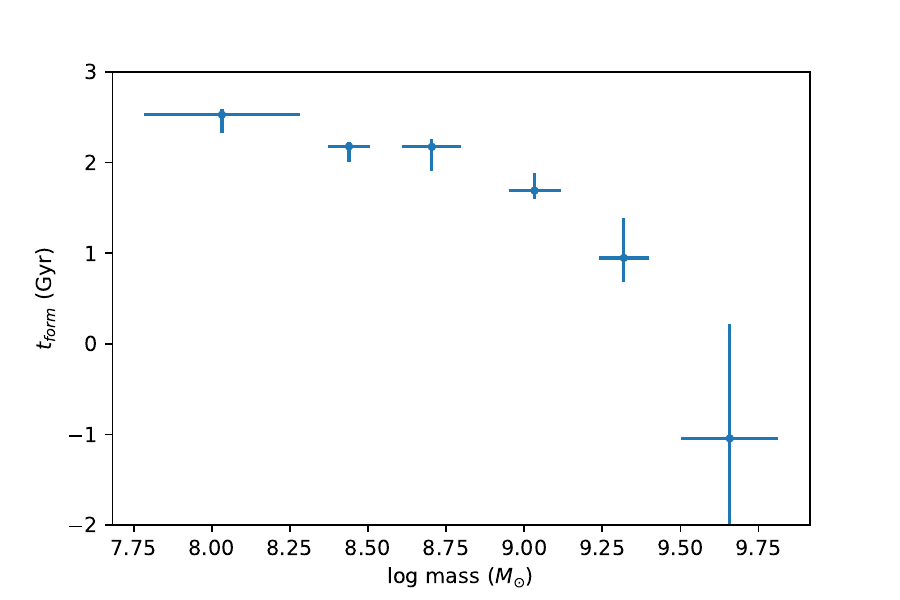}
\caption{The median formation time of the 200 galaxies vs. the galaxy stellar mass}
\label{fig:massage}
\end{figure}

 A similar analysis is carried out to investigate the correlation between the ages and stellar masses of galaxies. The 200 galaxies are divided into 6 bins according to their mass, with 33 or 34 galaxies in each bin. Then, the median age of each bin is calculated by the method mentioned above. The result is presented in Fig. 
\ref{fig:massage}. Also, a similar linear fit is carried out to test the relation between the stellar mass and age. The linear fitting result 
of the type $t=A+B\log M$ is $A=10.23_{-1.71}^{+1.92}$, $B=-0.95_{-0.19}^{+0.22}$, the slope $B<0$ within 10$\sigma$ confidence level, indicating a negative correlation between stellar mass and formation time $\rm(t_{form})$. This means that the more massive galaxies tend to be formed earlier, as our samples are not complete samples, we could not obtain a stronger conclusion.

\section{Discussion and Conclusion}
\label{sec:dis}

\cite{m2017} proposed a method to estimate the ages of galaxies by using two-SSP fitting, 
especially indicated for
SEDs with a V-shape typical of  massive galaxies with Lyman break and Balmer break, such as those found by \citet{labbe2023} at $z\gtrsim 6$. In this work, we applied the selection criteria of \citet{labbe2023} into the sources provided by the ZFOURGE catalog at redshifts between 2 and 4, excluding AGNs. We calculated the average ages of the galaxies in bin of different redshifts.

We obtained ages for 200 candidate galaxies, which we divided our galaxies into 6 reshift bins, and calculated the average ages for each bin by the method introduced in sect. \ref{sec:method}. They are compatible to be lower than the age of the Universe in the standard $\Lambda$CDM model. In addition, we only considered the candidates with rest R-band luminosity larger than $5.75\times 10^{42}$ $\mathrm{erg s^{-1}}$ (observable at any redshift, that is, free from Malmquist bias) and we got a similar result, which means that the luminosity selection does not have significant effects on our research. When we focused on the highest-redshift candidates in our samples, a similar result was obtained. In general, our average ages of the galaxies are between $0.52_{-0.29}^{+0.35}$ Gyr and $2.69_{-0.69}^{+0.94}$ Gyr (at redshifts 2.39 and 2.75, respectively).
Although the ages of galaxies obtained in our sample are roughly consistent with the age of the Universe, there are bins with significantly high average ages. For example, the fourth bin in Figure~\ref{fig:selection} has the average age of $2.69_{-0.64}^{+0.96}$ Gyr, while the corresponding cosmic age is 2.75 Gyr; the fourth bin in Fig. ~\ref{fig:agez} has the average age of $2.10_{-0.46}^{+0.60}$ Gyr, while the corresponding cosmic age is 2.80 Gyr. This means that the galaxies were formed when the age of the Universe was only 500 Myr. Also, in the paper, a negative correlation between the galaxies' stellar masses and formation time is obtained in our sample.
Note however that the systematic error caused by the two-SSP method may correspond to the irregular relation between the redshifts and ages obtained by our results \citep{m2017}. Some previous works have studied the ages of distant red galaxies in the redshift range between $z=1.2$ and $z=3.4$, and the obtained ages of candidates are within 1.7 Gyr to 5.5 Gyr \citep{Toft2005,Labbe2005}, which are consistent with our results, though the selection criteria may vary the result. Average ages larger than one Gyr are also obtained
if extremely red quiescent and massive galaxies at $2.5<z<3.8$ are selected \citep{Cas12,m2017}.

These ages of the oldest stellar population in galaxies are useful to constrain the cosmological model \citep{Trenti_2015,Jimenez_2019,valcin2020,bernal2021,bkm2021,wei2022}. In general, from both Figures \ref{fig:agez} and \ref{fig:selection}, we could see that the oldest galaxies in the range $2<z<4$ occur at redshift near 2, that is, the galaxies are younger at higher redshift,
as expected. However, we do not see a clear variation of the age of galaxies proportional to the variation of the age of the universe for $z>2.5$:
In Figure~\ref{fig:high}, we do not see a correlation of age with redshift, with all average ages of
galaxies remaining between 0.5 and 2 Gyr.

Since the launch of JWST, more and more massive galaxies in the high-$z$ Universe are being discovered. \citet{Lop24} have applied this methodology of two-SSP fitting to 13 JWST galaxies with the same SED characteristics at $\langle z\rangle=8$ and got an average age >0.9 Gyr (95\% CL), which presents a tension with the age of the Universe under $\Lambda $CDM cosmology, and
it is within the same range of ages than the galaxies at $2.5<z<4.0$. Within the samples selected by the color criterion of \cite{labbe2023}, there are no significant age evolution of their average or median ages between the redshifts, so it is reasonable to suppose that the ages of the double peak V-shaped galaxies do not have evolution in higher redshift, which means that if the galaxies with similar properties are found in high redshift (such as higher than 8), there may be a challenge to the present-day galaxy formation model.

\begin{acknowledgments}
The data used in this article were downloaded from URL
\url{https://cdsarc.cds.unistra.fr/viz-bin/cat/J/ApJ/830/51#/browse}.
ZFOURGE is a 45 night imaging program taken with the FourStar near-infrared camera on Magellan, which combines deep imaging in J1, J2, J3, Hs, Hl, and Ks with public legacy surveys to fit spectral energy distributions for $>60,000$ galaxies \citep{Straatman2016}.
M.L.C’s research is supported by the Chinese Academy of Sciences
President’s International Fellowship Initiative grant No. 2023VMB0001.
J.J.W’s research is supported by the Natural Science Foundation of China (grant No. 12373053), the Key Research Program of Frontier Sciences (grant No. ZDBS-LY-7014) of Chinese Academy of Sciences, and the Natural Science Foundation of Jiangsu Province (grant No. BK20221562).
\end{acknowledgments}

\bibliography{sample631}{}
\bibliographystyle{aasjournal}



\end{document}